# Valley Carrier Dynamics in Monolayer Molybdenum Disulphide from Helicity Resolved Ultrafast Pump-probe Spectroscopy

Qinsheng Wang[1,*], Shaofeng Ge[1,*], Xiao Li[1], Jun Qiu[1], Yanxin Ji[1], Ji Feng[1], Dong Sun[1, †]

[1]International Center for Quantum Materials, Peking University, Beijing 100871, China
[*]These authors contributed equally to the work.
[†]Email: sundong@pku.edu.cn

ABSTRACT

We investigate the valley related carrier dynamics in monolayer $MoS_2$ using helicity resolved non-degenerate ultrafast pump-probe spectroscopy at the vicinity of the high-symmetry K point under the temperature down to 78 K. Monolayer $MoS_2$ shows remarkable transient reflection signals, in stark contrast to bilayer and bulk $MoS_2$ due to the enhancement of many-body effect at reduced dimensionality. The helicity resolved ultrafast time-resolved result shows that the valley polarization is preserved for only several ps before scattering process makes it undistinguishable. We suggest that the dynamical degradation of valley polarization is attributable primarily to the exciton trapping by defect states in the exfoliated $MoS_2$ samples. Our experiment and a tight-binding model analysis also show that the perfect valley CD selectivity is fairly robust against disorder at the K point, but quickly decays from the high-symmetry point in the momentum space in the presence of disorder.

TABLE OF CONTENTS GRAPHIC

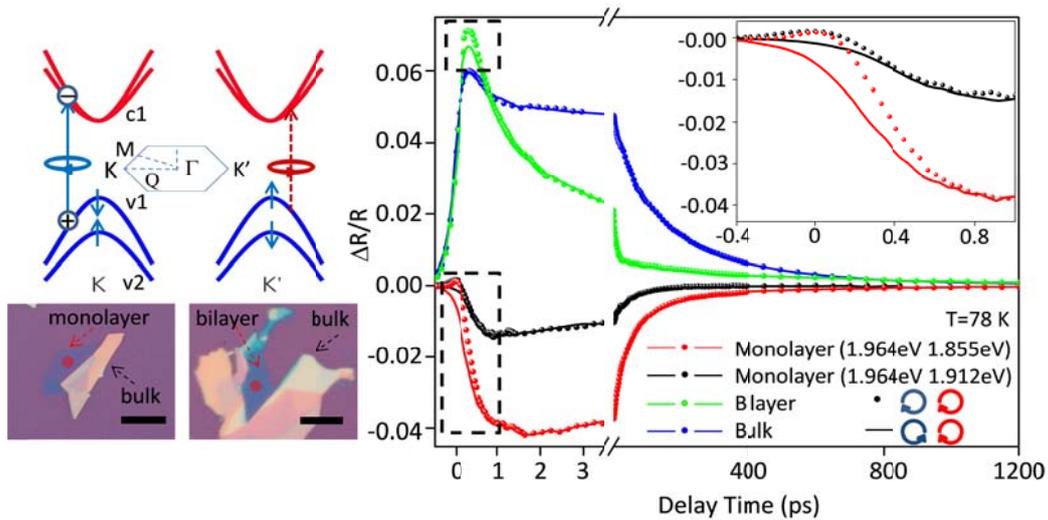

Molybdenum Disulphide (MoS$_2$) as a representative of the most explored two dimensional (2D) transition-meal dichalcogenides (TMDC), has been shown to exhibit remarkable optical properties during the past few years. These include the transition from indirect to direct bandgap when reaching monolayer from bulk,[1,2] trion state with large bonding energy,[3,4] efficient valley and spin control by optical helicity[5-7] and signature of coherence[8] between the two valleys due to its valley selective circular dichroism(CD).[5,9] These unique optical properties, as well as emergence of convenient and controllable growth method to acquire the monolayer materials, makes this category of materials especially interesting and promising for possible valleytronics applications.[10-13]

Toward the application of TMDC in future valleytronics, it is critical to examine the dynamics of incipient valley polarization by CD, and the subsequent relaxation dynamics, especially various scattering channels that can degrade the valley polarization. Previous photoluminescence (PL) measurement relies on the photon emission process to read out valley carrier distribution which only gives partial information about the valley degree of freedom even when time resolved PL measurement is performed.[14,15] This is because the scattering of carriers to non-radiative state cannot be detected in PL measurement. As a powerful tool for studying the various dynamics in materials and devices, ultrafast pump-probe spectroscopy has been applied to study thin layer and bulk MoS$_2$ recently.[16-19] However, existing experiments have not addressed the interesting valley dynamics due to two factors: 1) the lack of helicity resolution, and 2) large pump photon energy that is far away from the transition at the vicinity of K point. A helicity resolved ultrafast pump-probe experiment on monolayer MoS$_2$ with transitions at the vicinity of K point is highly desirable to resolve the valley related carrier dynamics, in the aims of both understanding the discrepancies in existing measurements: such as the variation of valley polarization from 30% to 100% in different experiments[5-7,20] and stepping toward future valleytronics applications: such as optical initialization and manipulation of the valley degree of freedom and coherence in TMDC.[8]

In this article, we present a helicity resolved optical non-degenerate pump-probe experiment on exfoliated monolayer MoS$_2$ in a reflection setup. Fig. 1a shows the lattice structure of MoS$_2$, which has a hexagonal crystal structure with covalently bonded S-Mo-S. Ab initio band structure results[21-25] (Fig. 1c) indicate that MoS$_2$ crosses over from an indirect gap semiconductor with multilayer to a direct gap semiconductor with monolayer. Because the two sublattices are occupied by one molybdenum and two sulphur atoms respectively, inversion symmetry is broken in monolayer MoS$_2$. The broken inversion symmetry, together with the broken spin degeneracy by spin-orbit coupling, leads to valley-dependent optical selection rules for interband transitions at K and K′ points: interband transitions at K (K′) valleys are allowed for optical excitation of left (right) circular polarized light only.[5,6,9]

As shown in Fig 1(d), valley polarized carriers are injected by left/right ($\sigma^-/\sigma^+$) circular polarized pump pulse and detected by measuring the differential reflection of left circular ($\sigma^-$) probe pulse. Except in pump wavelength dependent measurements, the pump and probe photon energies are centered at 1.964 eV and 1.912 eV respectively with 4 meV bandwidth, which is to be compared with the A-exciton energy[1,2] centered at 1.86 eV with 70 meV bandwidth from PL measurement (see supporting information). Both pump and probe photons excited transitions are from the highest valence band v1 to lowest conduction band c1 at the vicinity of K point (Fig 1c). The transition between spin-orbital interaction split valence band v2 and c1 is not directly excited by one photon absorption in a monolayer. Under perfect circular polarized valley selection,[5,9] the $\sigma^+$ pump photon excites carriers into K′ valley only, while the $\sigma^-$ probe photon is sensitive to pump induced carrier distribution change in K valley. Thus intervalley scattering of carriers between K and K′ can be dynamically measured by varying the pump and probe photon delays. Here we measure the change of the probe pulse reflection, ΔR, induced by the pump. ΔR is then normalized by the probe reflection to obtain the differential reflection, ΔR/R= (R′-R)/R, where R′ and R are the reflection of the probe pulse from the sample with and without the presence of the pump pulse, respectively.

**RESULTS AND DISCUSSION**

Figure 2 shows a representative helicity resolved transient reflection spectra of monolayer, bilayer and bulk $MoS_2$ at 78 K. The monolayer shows negative ΔR except at the vicinity of time zero. In contrast, both bilayer and bulk show positive ΔR during the entire course of the measurements. A remarkable observation is that the initial transient reflection spectra of monolayer are different with $\sigma^+$ and $\sigma^-$ pump excitations. The difference is observable over period of about 1 ps. This difference varies on different places of one sample and on different samples. A large variation from 1 to 7 ps is observed on different samples (see supporting information). In contrast, the transient reflection of bulk is independent of pump helicity. If the pump photon energy increases slightly to 2.072eV (moving the pump photon excitations by 212 meV away from the K point), the pump helicity dependence completely disappears (Fig. 2b). The temperature dependent measurements show the pump helicity dependence of ΔR on monolayer persists up to 298 K (Fig 3a). In contrast to the full negative ΔR over entire delays for $\sigma^+$ pump excitation at 78 K, the ΔR turns positive at the vicinity of time zero at 298 K. Fig 3b shows the positive ΔR component also increases with temperature for $\sigma^-$ pump excitation. The pump helicity dependence of transient reflection is also observed in bilayer due to the broken inversion symmetry by coupling to the substrate,[6,26] which disappears at room temperature in as-prepared sample (see supporting information).

The decay from the negative peak of ΔR shown in Fig 3c is multi-exponential with two distinct time scales, namely a fast component ($\tau_1$) of 7.1 ps and a slow component ($\tau_2$) of 61.3 ps at 78 K. Both $\tau_1$ and $\tau_2$ increase with temperature, indicating a phonon related process during both decays (Fig. 3d). According to the previous studies,[14,16] we attribute the fast decay to phonon scattering and the slow decay to nonradiative interband electron hole recombination. The slow component is much faster in monolayer than in bulk due to the transition from direct bandgap to indirect bandgap.[1,2,16]

We now turn to the interpretation of the sign of the transient reflection signal observed in the experiment. The positive ΔR of bulk and bilayer is consistent with the free carriers or excitons state filling effect as studied previously in the literature.[16-18] The pump induced free carriers or excitons occupy the probe transition states and reduce the probe photon absorption. Thus the Pauli blocking enhances the probe reflection, leading to a positive ΔR. As carrier occupation effect dominates ΔR at moderate carrier excitation intensities, many-body effects become significant in semiconductors with reduced dimensionalities or at high carrier excitation intensities.[27-33] In the case of highly excited (typical pump photon fluence of $4*10^{14}/cm^2$) two-dimensional monolayer $MoS_2$, the contribution of many-body effects to ΔR cannot be neglected, though the initial positive ΔR can be dominated by an instantaneous exciton ground state bleaching.[34] Because we use probe photon energy equivalent to the exciton transition at the band edge, the band gap renormalization effect due to the coulomb interactions of the dense electron-hole plasma contributes to the negative ΔR signal[28,30-33] and matches the sign of ΔR in monolayer that we measured. Here we rule out the possibility that the negative sign of ΔR of monolayer is from a resonant transition of probe as observed by similar experiment in GaAs,[35] since the probe photon energy is close to the resonance of A-exciton in bilayer and bulk too. The monolayer sample also shows the same ΔR sign when we switch the probe wavelength to non-resonant 633 nm (see supporting information), which further rules out the resonant transition effect. The argument that the bandgap renormalization may dominate the ΔR signal is also supported by the decrease of positive ΔR at the vicinity of time zero with decreasing temperature as measured in Fig 3b, which matches temperature dependence of the bandgap renormalization effect.[28,32]

As to the helicity resolved results on monolayer, the independence of pump helicity when pumping with 600 nm is in good agreement with previous PL excitation measurement.[15,36] However, from our dynamics measurement, an initial valley carrier polarization by the excitation is not observed within the ~200 fs time resolution (Fig. 2b). Thus, the absence of the valley polarized CD is not attributable to the phonon assisted intervalley scattering when carriers relax from relatively high excited states,[36] as carrier phonon scattering timescale is larger than 200-fs. The only interpretation would be that the valley CD selectivity decays quickly when the excitation is away from the vicinity of the high symmetry K point, and thus 600 nm σ⁻ pump excites carriers almost equally in K and K′ valleys initially. For a perfect crystal without disorder, theoretical calculations predict nearly unity valley CD selectivity over almost the whole K and K′ valleys.[5,9] However, in an exfoliated $MoS_2$ sample with significant disorder, the valley CD selectivity decays due to the presence of disordered defects. A tight binding simulation of valley CD selectivity as a function of disorder strength W, which is an additive onsite energy perturbation akin to Anderson disorder, is shown in Fig 4 (simulation details in supporting information). When there is no disorder (*W*= 0 eV), the valley CD selectivity at K is 1. We choose a small region around K in which the valley CD selectivity is larger than 98.5%. Upon the introduction of disorder, the valley CD selectivity decreases. The valley CD selectivity at K decreases to 80% when W = 0.6 eV, while valley CD selectivity away from K point decreases, by a much greater amount, down to 20% (Fig. 4b). Therefore, disorder has pronounced effect on the valley CD selectivity for off-resonance transitions, but the on-resonance transition at the high symmetry K point is fairly robust against disorder. As a result, when pumping closer to the K point at 633 nm, the valley polarization can still be created.

The valley polarization excited by 633nm only preserves for several ps before the scattering process makes the carriers distribution in K and K′ valley equivalent. We attribute the dominant mechanism of this fast scattering to the trapping of exciton by defects. The trapped exciton (or localized exciton) has been observed to emit at 1.8 eV with no valley CD selectivity.[6,15] For monolayer with significant surface-to-volume ratios, the surface defects can act as exciton traps, and this trapping process can be very efficient.[37] The efficient exciton trapping can account for the very short A-exciton emission life time[14,15] and low PL quantum yield observed on monolayer $MoS_2$.[6,8,14,15] In previous ultrafast differential transmission measurement on suspended $MoS_2$ sample and time resolved PL measurement,[14-16] the scattering time is observed to be 2~4 ps. This matches the several ps scattering time measured in our experiment considering it can varies from sample to sample with different disorder strengths. Here we can not fully rule out the possibility that optical phonon scattering can contribute to the intervalley scattering process considering its probable ps time scale. However, we infer an optical phonon scattering dominated intervalley scattering process is unlikely for at least two reasons: first, optical phonon scattering may have stronger temperature dependence compared to the weak temperature dependence from 78 K to 298 K observed in our experiment (Fig. 3a). Second, the intervalley phonon scattering time should not have large variation on different samples, especially on different spots on one sample as observed in sample dependent measurement (supporting information).

At last, we want to emphasize that the fast exciton trapping process does not degrade the valley CD selectivity observed in a helicity resolved steady state PL measurement. The excitons trapped by disorder are not counted due to its low radiative efficiency and longer emission wavelength in a PL measurement, so defect scattering to trapped exciton is not an intervalley scattering process, although it degrades the PL emission efficiency at measured wavelength. The defect scattering time is attributed to the exciton decay time $\tau_A$ of the valley CD selectivity measured in steady state PL: $\rho=1/(1+2\tau_A/\tau_{AS})$ (equation 1 in Ref. 6). Intervalley scattering process (for example, intervalley phonon scattering) to a bright state in the other valley, however, can degrade the valley CD selectivity measured in PL and is attributed to the intervalley scattering time of $\tau_{AS}$. In a defect scattering dominated sample (which is true in most current exfoliated natural flake of $MoS_2$ that is abundant with defects), the relation $\tau_A \ll \tau_{AS}$ holds, and thus $\rho \sim 1$.

The above equation of valley polarization assumes unity valley selection during the excitation. If we further consider the valley initialization is not unity due to the presence of disorder strength under slightly off resonance excitation, the valley CD selectivity measured in a steady state PL measurement does correlate with the disorder strength during optical valley polarization initialization stage and strongly depends on pumping photon energy at large disorder strength as observed in the literatures.[15,36] In the limit that defect scattering time is much faster than the intervalley scattering time, the degree of valley polarization that is measured in a steady state PL is mainly determined by the valley polarization during the valley carrier initialization process. Lower degree of valley polarization from PL usually indicates larger disorder strength in the sample, which can potentially enhance the defect scattering rate and thus give shorter carrier valley life time in transient measurement. However, we want to clarify that different species of disorders or defects can play different roles in these processes. They may affect the valley polarization initialization or exciton trapping in different manners. These factors will complicate the relationship between transient measurement and steady state PL, which requires further studies.

**CONCLUSION**
In the prospects of valleytronics application of TMDC, high quality sample with lower disorder strength is critical in two respects: suppressing the decay of valley CD selectivity of excitation away from K point to get better optical initialization of valley polarization; and suppressing the disorder dominated scattering to keep longer valley polarization for manipulating valley carriers. At current stage, exfoliated monolayer WSe$_2$ shows much better quantum yield and valley polarized CD persists at excitation energy far away from vicinity of K point.[8] This indicates much lower disorder strength compared to MoS$_2$. Improved sample quality can be further achieved by putting TMDC on flat substrate such as BN[6] and using improved growth method of monolayer TMDC for large scale production, such as chemical vapor deposition[10] and molecular beam epitaxy. Similar dynamics measurements on such a high quality samples can potentially provide information of the valley polarization life time limited by intervalley phonon scattering.

In summary, our pump-probe measurement elucidates the probable important role of many-body effects in the transient reflection signal of monolayer MoS$_2$ due to reduced dimensionality. Helicity resolved ultrafast dynamics measurement indicates that the disorder scattering is the dominant valley polarization relaxation channel in exfoliated monolayer MoS$_2$. In the presence of disorder, the valley CD selectivity relaxes quickly when excited away from the high-symmetry K and K′ valley. Thus, a resonant optical excitation is essential for valley polarization initialization in highly disordered sample. Future improvement on sample quality with low disorder strength is desired for claimed valleytronics application of TMDC.

**Methods**

**Sample preparation**. Monolayer MoS$_2$ flakes are obtained by mechanical exfoliation of a natural bulk MoS$_2$ crystal on a Si/ 285nm SiO$_2$ substrate. The number of layers is identified by optical contrast in a microscope initially and confirmed by micro-Photoluminescence spectroscopy and micro-Raman Spectroscopy as shown in supporting information.

**Helicity resolved non-degenerate ultrafast pump probe spectroscopy**. In the non-degenerate ultrafast pump probe setup, a 60-fs 250-kHz amplified Ti: sapphire laser at 800 nm is separated into two arms, both focused on a 2-mm sapphire plate to generate white light supercontinuum. Then a narrow bandpass filter is used to filter the desired pump and probe wavelengths for the pump probe measurement. Both pump and probe photons are cross or co-linear polarized before entering a Fresnel-Rhomb to get cross or co-circular polarized light, which is subsequently focused onto the sample that's placed in liquid nitrogen cooled cryostat for temperature control through a 40X near infrared objective. Due to the tight depth of the focus of the high numerical aperture objective and 285 nm thick SiO$_2$ on silicon, the background pump probe response from Si substrate is negligible as measured on substrate with no MoS$_2$ (see supporting information).The temporal resolution is about 200 fs, determined through both cross correlation measurements and ΔR rise time. The reflected probe is collected by the same objective lens and detected by a Si photodetector after depolarizer and lock-in amplifier referenced to 5.7-kHz mechanically chopped pump. A narrow band filter at probe wavelength is used to prevent the detection of reflected pump. An additional notch filter center at pump wavelength is added when the pump probe wavelength are close to each other. The probe spot size is estimated to be below 2 μm with pump spot size slightly larger depending on the pump wavelength. The pump fluences are kept around $4*10^{14}$ photons/cm$^2$ with pump wavelength of 633 nm (1.964 eV). Pump power dependent dynamics measurement shows linear dependent with pump fluence and this rules out the possibility that higher order effect may dominate the transient response (see supporting information).

**Supporting Information Available**: Sample photoluminescence and Raman characterization, additional helicity resolved ultrafast spectroscopy figures and computational details. This material is available free of charge via the internet at http:// pubs.acs.org.

**Acknowledgement**:

The authors want to acknowledge Theodore B. Norris and Qian Niu for helpful discussion, B.L. Liu and P. H. Tan for providing exfoliated $MoS_2$ sample at the early stage of our experiment. This project has been supported by the National Basic Research Program of China (973 Grant Nos. 2012CB921300 and 2013CB921900), the National Natural Science Foundation of China (NSFC Grant Nos. 11274015 and 11174009) and the Specialized Research Fund for the Doctoral Program of Higher Education of China (Grant No.20120001110066). All correspondence and requests for materials should be addressed to D. Sun.

**Figure Captions:**

Figure 1. (a) Lattice structure of $MoS_2$ (b) Optical micrograph of $MoS_2$ on 285 nm $SiO_2$. The red spot marks the probe position in spectroscopy measurement. Scale bar: 10 μm. (c) Band diagram of monolayer (solid) and bulk (dash) $MoS_2$ and pump (blue) probe (red) photon transitions configuration. (d) The lowest-energy conduction bands and the highest-energy valence bands, with valley selection rule for circular polarized light.

Figure 2. (a) Helicity resolved transient reflection dynamics at 78 K on monolayer (black and red), bilayer (green) and bulk $MoS_2$ (blue). The pump photon is set at 1.964 eV (633 nm) for all curves in the figures. The probe photon energy is set at 1.855 eV (650 nm) for comparison of transient reflection dynamics of samples with different thickness. For monolayer, probe photon energy of 1.912 eV (670 nm) is measured for comparison. The inset shows the signal at the vicinity of time zero of monolayer. (b) Helicity resolved transient reflection dynamics of monolayer $MoS_2$ with pump photon energy of 2.072 eV (600nm) and probe photon energy of 1.912 eV (650nm) at 77 K. The lines and dots show results from cross and co-circular pump probe polarization configurations in all figures.

Figure 3. Temperature dependence of helicity resolved transient reflection dynamics of monolayer $MoS_2$ with pump wavelength of 633 nm and probe wavelength of 650 nm. (a) Transient reflection dynamics of monolayer at 78 K (red) and 298 K (black) with cross- (line) and co-circular (dot) polarized pump probe configurations. (b) Evolution of transient reflection dynamics at vicinity of time zero with temperature from 298 K to 78 K measured with co-circular pump probe polarization configuration. (c) Two exponential decay function of $A \cdot \exp(-t/\tau_1) + B \cdot \exp(-t/\tau_2)$ fitting of transient reflection decay tails from negative peak at different temperatures. (d) Fitting parameters, fast ($\tau_1$) and slow ($\tau_2$) decay time constant for decay curves in Fig. c.

Figure 4. (a) Left: The band structure of $MoS_2$ monolayer along Γ→K from density functional theory (DFT) calculations. There is a direct gap of 1.86 eV at K point from the simulation. Right: Zooming into the region bounded by dashed lines in left panel. The bands based on DFT are shown by solid line and the ones from the tight-binding (TB) calculation are shown by dashed line. The vertical arrows with different colors represent the optical transitions, where the transition energy is 1.860 eV, 1.885 eV, 1.910 eV, 1.935 eV and 1.960 eV as we move away from K, respectively. (b) The optical selectivity (η) decays with disorder (W). Every colored line corresponds to the transitions with the same color in (a). The numbers near lines show the transition energy relative to the direct band gap.

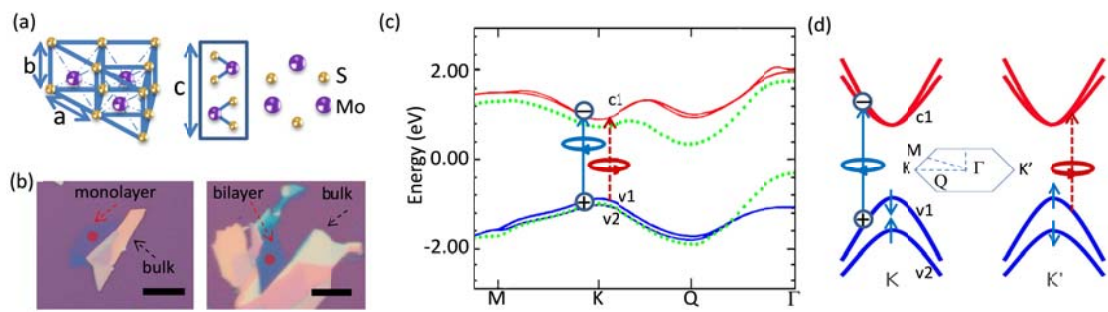

Figure 1

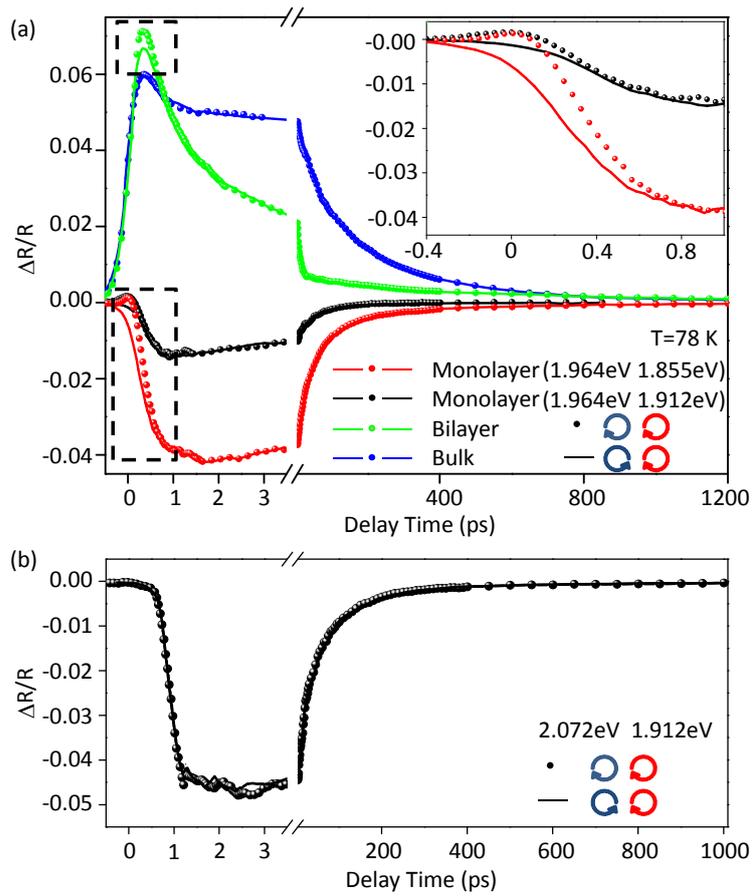

Figure 2

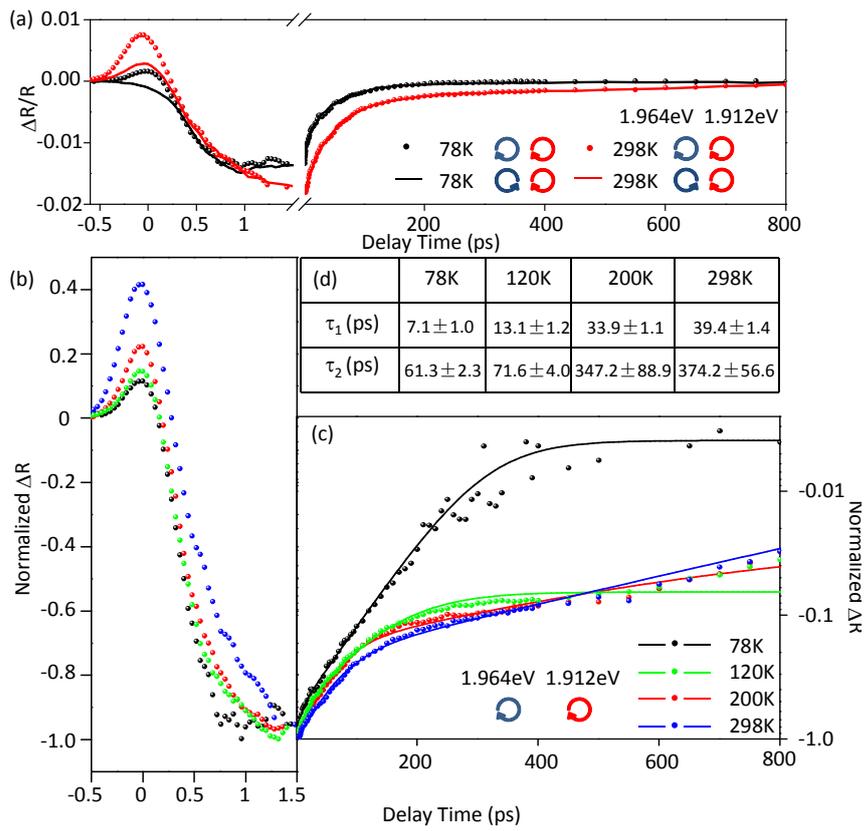

Figure 3

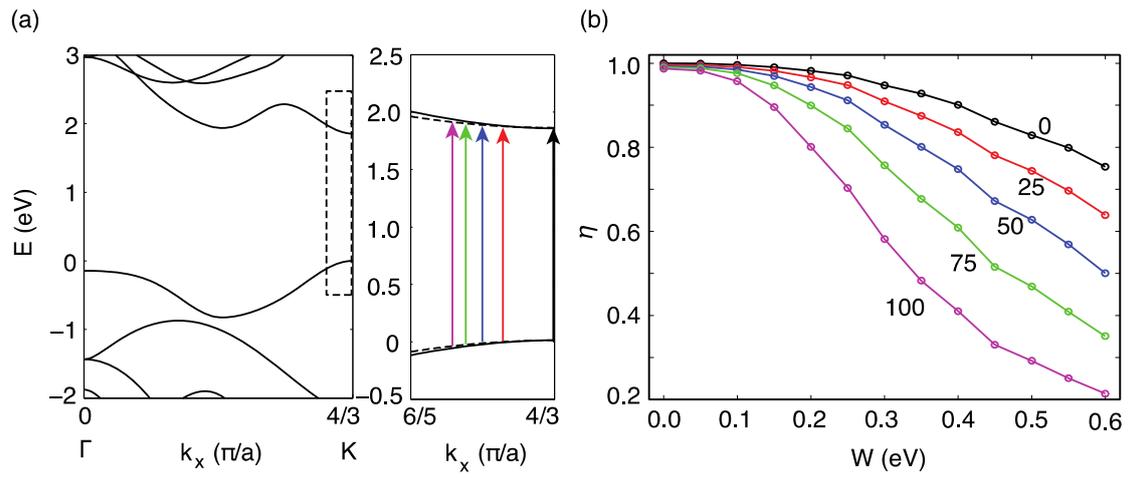

Figure 4